\input epsf
\documentstyle[preprint,aps,eqsecnum,tighten]{revtex}
\preprint{\vbox{\baselineskip=12pt
\rightline{CGPG-96/6-1}
\rightline{hep-th/9701136}}}
\def\be{\nopagebreak[3]\begin{equation}}
\def\ee{\end{equation}}
\def\ba{\nopagebreak[3]\begin{eqnarray}}
\def\ea{\end{eqnarray}}
\def\nl{\nonumber \\}

\def\d{{\rm d}}
\def\GL{{\cal GL}}
\def\x{\bar x}
\def\y{\bar y}
\def\s{\bar s}
\def\k{\bar k}
\def\H{{\cal H}}
\def\F{{\cal F}}

\newcommand{\teta}{\rlap{\lower2ex\hbox{$\,\tilde{}$}}\eta{}}
\begin{document}
\draft
\title{Gauss Linking Number and \\
Electro-magnetic Uncertainty Principle}
\author {Abhay Ashtekar\thanks{Electronic 
address: ashtekar@phys.psu.edu}
and Alejandro Corichi\thanks{Electronic 
address: corichi@phys.psu.edu}
}
\address{Center for Gravitational Physics and Geometry \\
Physics Department, Penn State  \\
University Park, PA 16802, U.S.A.}
\maketitle

\begin{abstract}

It is shown that there is a precise sense in which the Heisenberg
uncertainty between fluxes of electric and magnetic fields through
finite surfaces is given by (one-half $\hbar$ times) the Gauss linking
number of the loops that bound these surfaces. To regularize the
relevant operators, one is naturally led to assign a framing to each
loop.  The uncertainty between the fluxes of electric and magnetic
fields through a single surface is then given by the self-linking
number of the framed loop which bounds the surface.

\end{abstract}


\section{Introduction}

In 1833, Gauss noticed a striking fact about electro-magnetism \cite{1}.
He considered a loop $L_1$ carrying a constant current $I$ and
computed the work $W$ done in moving a magnetic monopole of strength
$m$ along a closed path $L_2$ in the magnetic field produced by the
current:
\be
W =  \frac{mI}{4\pi}\,\, \oint _{L_1}\d s \oint_{L_2}
\d t\,\,\epsilon_{abc} \,\dot{L_1}^a(s)\,
\dot{L_2}^b(t) \frac{L_1^c(s)- L_2^c(t)}{|L_1(s) - L_2(t)|^3}\, .
\label{g}
\ee
He then made a deep observation which can be stated in the modern
mathematical terminology as follows: although the double integral
\be
\GL(L_1, L_2)  :=  \frac{1}{mI}\, W
\label{glno}
\ee
makes use of Euclidean geometry in several ways, its value is in fact
a topological invariant, a measure of the linking between the loops
$L_1$ and $L_2$.  In particular, even if one deforms the loops, the
value of the double integral does not change so long as the loops do
not touch or cross each other.  This is a remarkable property and
Gauss expressed the belief that the quantity $GL(L_1, L_2)$ may have a
fundamental significance.  The view was shared by others.  In
particular, in his celebrated treatise on electricity and magnetism,
Maxwell returns to this property and further elaborates on it
\cite{2,5}.
 
It turns out that the double integral $\GL(L_1, L_2)$ does have a
fundamental significance in electro-magnetism, which however, could not
have been guessed before the advent of quantum field theory.  To see
this, consider source-free Maxwell theory (in Minkowski space-time).
In the Hamiltonian treatment, the vector potential $A_a(\x)$ and the
electric field $E^a(\x)$ (on a constant time hyper-plane) serve as the
basic canonically conjugate fields with the Poisson bracket relations:
\be
\{ A_a(\x), \, E^b(\y)\}  =  \delta_a^b\, \delta^3(\x,\y).
\label{ss}
\ee
(Throughout this paper, curly brackets will denote Poisson brackets.)
The vector potential itself is not an observable since it fails to be
gauge invariant.  However, we can integrate it over a closed loop
$\alpha$ to obtain a gauge invariant functional:
\be
B[\alpha] := \oint_\alpha A_a \d l^a \equiv
\int_{S_\alpha} {B}^a \d ^2{S}_a
\ee
where $S_\alpha$ is {\it any} 2-surface bounded by the loop $\alpha$.
Similarly, given any 2-surface $S_\beta$ bounded by a closed loop
$\beta$ we can define the flux of the electric field:
\be 
E[\beta] := \int_{S_\beta} {E}^a \d ^2{S}_a
\ee
which depends only on the loop $\beta$ (and not on the specific
surface $S_\beta$ with boundary $\beta$) because ${E}^a$ is
divergence-free.  The observables $B[\alpha]$ and $E[\beta]$ are
(over)complete in the sense that their values at a point $(B^a, E^a)$
of the physical phase space suffice to determine that point uniquely.

It is straightforward to compute the Poisson brackets between these
observables if $\alpha$ and $\beta$ have no point in common. The
result is:
\ba
\{ B[\alpha],  E[\beta] \}  &= &  \oint_\alpha \d l^a(\x) 
\int_{S_\beta} \d^2S_a(\y) \, \delta^3(\x, \y)\nonumber\\
&=& I(\alpha, \, S_\beta) \label{ino}
\ea
where $I(\alpha, S_\beta)$ denotes the oriented intersection number
between the loop $\alpha$ and the surface $S_\beta$.  But, as a
geometrical picture makes it clear (see figure 1), this intersection
number is precisely the linking number between loops $\alpha$ and
$\beta$.  (An analytic calculation showing the equality of $I (\alpha,
S_\beta)$ with $\GL(\alpha,\beta)$ is given in the Appendix.) Thus,
we have:
\be
\{ B[\alpha], E[\beta] \}  =  \GL(\alpha, \beta)
\label{pb}
\ee
The fact that the Poisson bracket is metric independent may seem
surprising at first. But note that, since the vector potential $A_a$
is a 1-form and the electric field $E^a$ (being canonically conjugate
to $A_a$) is naturally a vector density of weight one, neither the
symplectic structure (\ref{ss}) nor the definitions of the observables
$E[\alpha]$, $B[\beta]$ require a metric (or any other background
field) for their definitions.  Hence, if well-defined, the right side
of (\ref{pb}) {\it has to be} a topological invariant of loops
labeling the observables.

\bigskip

\centerline{
\hfill\vbox{\epsfxsize=5in\epsfysize=2.1in\epsfbox{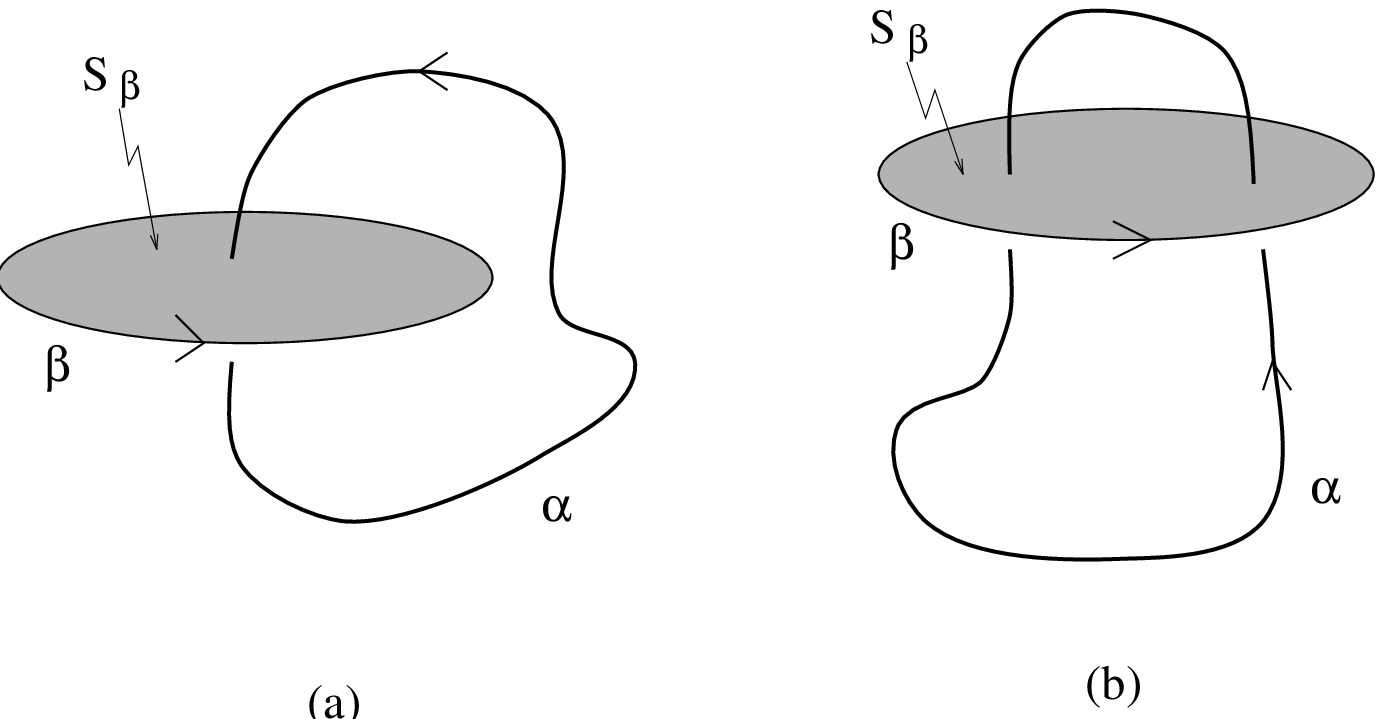}}\hfill}
\bigskip

\noindent
{\small {\bf Fig. 1}
The Gauss linking number $\GL(\alpha, \beta)$ equals the oriented 
intersection number $I(\alpha, S_\beta)$. In the case (a) this number
is $1$ and in (b) it is $0$.}  
\bigskip

Let us pass to the quantum theory heuristically. One would expect that
(if $\alpha$ and $\beta$ have no point in common), the commutator
between the magnetic and electric flux operators would be given by:
\be
\Big[\hat{B}[\alpha], \,\hat{E}[\beta] \Big] = 
{i\hbar}\,\GL(\alpha, \beta) 
\label{ccr}
\ee
and hence the Heisenberg uncertainties should satisfy:
\be
(\Delta \hat{B}[\alpha]) (\Delta \hat{E}[\beta]) \ge 
\frac{\hbar}{2} \, \GL(\alpha, \beta)\, .  \label{ur}
\ee
This implies that there is an intrinsic uncertainty in the
simultaneous measurements of fluxes of electric and magnetic fields
across finite surfaces if the Gauss linking numbers of the loops
bounding the two surfaces is not zero (see figure 2). In this sense,
as suspected by Gauss and others,the linking number does have a
fundamental significance in electro-magnetism.
\bigskip

\centerline{
\hfill\vbox{\epsfxsize=3.5in\epsfysize=1.5in\epsfbox{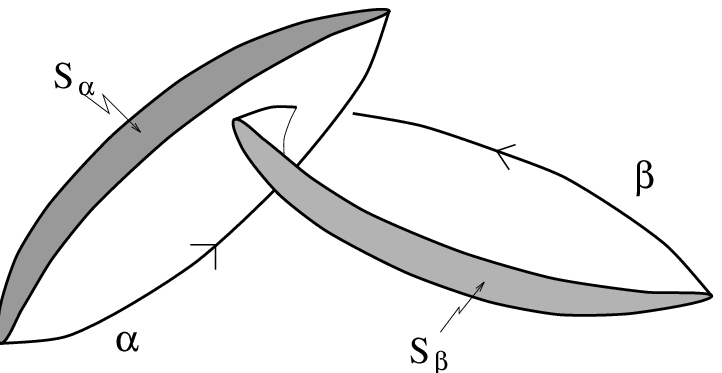}}\hfill}
\bigskip

\noindent
{\small {\bf Fig. 2} The (absolute value of the) Gauss linking number
between loops $\alpha$ and $\beta$ is $1$. In this case, the
Heisenberg uncertainty between the magnetic flux through the surface
$S_\alpha$ and the electric flux through $S_\beta$ is $\frac{\hbar}{2}$. 
\bigskip

A number of questions arise immediately.  Can one make these quantum
considerations precise?  If the quantization is {\it based} on the
algebra of operators generated by $\hat{B}[\alpha]$ and
$\hat{E}[\beta]$, the answer is clearly in the affirmative. However,
in such representations, the Maxwell Hamiltonian operator fails to be
well-defined. In the standard Fock representation where the
Hamiltonian {\it is} well-defined, operators $\hat{E}[\alpha]$ and
$\hat{B}[\beta]$ fail to be well-defined! (See, e.g., [5].)  Can one
nonetheless give meaning to the topological uncertainty relations
(\ref{ur}) in the Fock representation {\it in a suitable limiting
sense}?  Secondly, the uncertainty relation given above fails to be
meaningful if the loops $\alpha$ and $\beta$ coincide.  However, it is
known that the Gauss self-linking number of a loop {\it is}
well-defined if the loop is framed \cite{5}. Is there perhaps a
framing that is naturally introduced in the process of regularization
of the operators in question?  If so, can one give meaning to the
commutator of $\hat{B}[\alpha]$ and $\hat{E}[\alpha]$?  The purpose of
this note is to analyze these issues.  We will find that the specific
questions raised here can be answered affirmatively.

In Sec. II, we will consider the Fock space of photons and show that
suitably `thickened', regulated versions of the above `flux operators'
are indeed well-defined on the Fock space. This will enable us to
regard $\hat{B}[\alpha]$, and $\hat{E}[\beta]$ as certain limits of
well-defined operators.  We will see that the commutation relations
(\ref{ccr}) also holds in a limiting sense.  In this limit, the
thickening of surfaces goes to zero. However, the loop that bounds the
limiting surface carries the `memory' of the thickening in the form of
a framing.  We will see in Sec. III that the limits of the
commutators of the regulated flux operators are functionals of these
framed loops.  In particular, the framing enables one to evaluate the
(limits of) commutators without ambiguities even when the loops
intersect and overlap. We should emphasize however that the level of
rigor in this note is that generally used in theoretical physics
rather than mathematical physics.

We will conclude this section with a few remarks.\hfil\break
i) One often defines {\it dimension-less} observables: $B'[\alpha]=
\frac{c}{e}\, B[\alpha]$ and $E'[\beta] = \frac{1}{e}\, E[\beta]$,
which can be exponentiated to obtain Weyl commutation relations. (The
exponential of $iB'[\alpha] \equiv i\frac{c}{e}\oint_\alpha
\d l^a A_a$, for example, is the $U(1)$ holonomy.) In terms of these
primed observables, the uncertainty principle reads: $(\Delta
B'[\alpha])(\Delta E'[\beta]) \ge (1/2\alpha_{\rm fine})
\GL(\alpha, \beta)$, where $\alpha_{\rm fine}$ is the fine structure 
constant.

ii) In the non-Abelian case, one can replace $\exp\, iB'[\alpha]$ by
the trace of the holonomy of the connection along the loop
$\alpha$. The analog of $E'[\beta]$ is trickier. To ensure gauge
invariance, one now has thicken the loop to a ribbon. (See, e.g., 
\cite{3,9}.) The commutator is then again `topological'. However, the 
physical meaning of these observables is now less transparent.

iii)Over the last five years, the uncertainty relation (\ref{ur}) was
used by one of us (AA) to motivate the loop representation for gauge
theories and gravity in a number of conferences (see e.g. \cite{3}).
It was pointed out that a similar observation on the Maxwell
uncertainties was made by J\"urgen L\"offelholz from the Leipzig
mathematical physics group and also by condensed matter theorists
interested in flux quantization. Unfortunately, however, we have not
been able to find specific references.

\section{ Regulated  Flux Operators}
\subsection{Preliminaries}

Let us begin by recalling a few facts about the Fock representation of
photons.  Since we are interested in (fluxes of) electric and magnetic
operators, it will be convenient to adapt the discussion to a
canonical framework.  A vector $V$ in the 1-photon Hilbert space $\H$
is then represented by a pair $(A_a, E^a)$ of divergence-free vector
fields on a constant time hyper-plane and the Hilbert space norm is
given by (see, e.g., \cite{4}):
\be
\langle V|V\rangle= \frac{1}{\hbar}\,
\int_{\Sigma} \d^3x\,\left[A_a({\triangle}^{1/2}A^a) +
 E^a({\triangle}^{-1/2}E_a)
\right] \, .
\ee
Denote by $\F_\H$ the symmetric Fock space based on $\H$.  Electric
and magnetic fields are represented by operator valued distributions
involving the standard linear combinations of creation and
annihilation operators on $\F_\H$.  Consider, for example, the smeared
object,
\be
\hat{E}[f] :=  \int \d^3x \, \hat{E}^a (x)f_a(x)
\ee
where $f_a$ is a test-field, i.e., vector field of compact support.
Since $E^a$ is divergence-free, we have: $E[f] = E[f+\partial g]$, for
any test function $g$.  This is a well-defined operator on $\F_\H$
provided the vector $V = ({}^T\!f_a, 0)$ lies in the Hilbert space $\H$,
i.e., provided the norm
\be
\langle V|V\rangle=\frac{1}{\hbar}\int  \d^3x\,\left[{}^T\!\!f_a\,
({\triangle}^{1/2}\,\,\, {}^T\!\! f^a)\right] 
\equiv \frac{1}{\hbar}\,\int \d^3k |k|\,\, |{}^T\!\tilde{f}_a|^2 \label{norm}
\ee
is finite, where ${}^T\!\!f_a$ is the transverse part of $f$, and
${}^T\!\tilde{f}$, its Fourier transform. (The transverse projection
removes the `gauge freedom' of adding a gradient to $f_a$.)  In that
case, $\hat{E}[f]$ is expressible as a sum of the creation and
annihilation operators associated with the state $V$.  

The situation with the magnetic field operator is completely
analogous. The commutator between smeared electric and magnetic fields
is given by:
\ba
\Big[\hat{B}[f]\,, \hat{E}[g] \Big] &=& i\hbar\, 
\int \!\d^3x\,\epsilon^{abc}
(\partial_a f_b(\x)) g_c(\x) \nonumber\\
&=& \hbar\, \int \!\d^3k \epsilon^{abc}k_a (\tilde{f}_b (\k))^\star
\, \tilde{g}_c(\k) \label{scr}
\ea 
where $\tilde{f}_b$ denotes the Fourier transform of $f_b$ and $\star$
denotes complex conjugation.

\subsection{Regularization}

Let us now consider the formal expression of the electric flux
operator:
\be
\hat{E}[\beta] = \int_{S_\beta} \hat{E}^a  \d^2S_a \, .
\ee
It can be expressed as a smeared electric field, $E[\beta] = \int
\d^3x E^a (\x)f^{(\beta)}_a (\x)$, where, however, the test field
$f^{(\beta)}_a(\x)$ is a {\it distribution} with support on $S_\beta$:
\be
f_a^{(\beta)} (\x)  = \int_{S_\beta}\! \d^2S_a \,\,
\delta^3 (\x , \s_\beta)\, ,
\ee
where $\s_\beta$ denotes a point on the surface $S_\beta$.  Hence, the
corresponding operator $\hat{E}[\beta]$ is only formal; it fails to be
well-defined on the Fock space.  We must regulate it.

We will proceed in two steps (the first of which is the crucial one).
Geometrically, the problem arises because $f^{(\beta)}_a$ is a
distribution with two dimensional support.  We can remedy this
situation by an appropriate `thickening' of the surface $S_\beta$.
Let us therefore replace the loop $\beta$ by a strip (or ribbon)
$\Sigma_\beta$ of height $\epsilon$ (see figure 3).  More precisely,
let us proceed as follows. Let us first equip $\beta$ with a framing
(i.e., let us introduce, at each point of $\beta$, a vector in a
direction transverse to $\dot\beta^a$, the tangent to $\beta$.)  Then,
for each $\tau \in [0, \epsilon]$, let us denote by $\beta_\tau$ the
loop obtained by displacing $\beta$ a distance $\tau$ along the
framing. (Thus $\beta_0 \equiv \beta$). This construction uses the flat
Euclidean metric on the spatial hyper-plane. But the key final results
will not depend on this flat metric.)  Let $S_\tau$ denote a surface
bounded by the loop $\beta_\tau$ (such that the assignment $\tau
\rightarrow S_\tau$ is smooth.) The family of loops
$\beta_\tau$ constitute the strip $\Sigma_\beta$ and the
three-dimensional region swept out by the family of surfaces $S_\tau$
constitutes a `pill-box' $P_\beta$ with boundary $\Sigma_\beta$.
\bigskip

\centerline{
\hfill\vbox{\epsfxsize=3in\epsfysize=2in\epsfbox{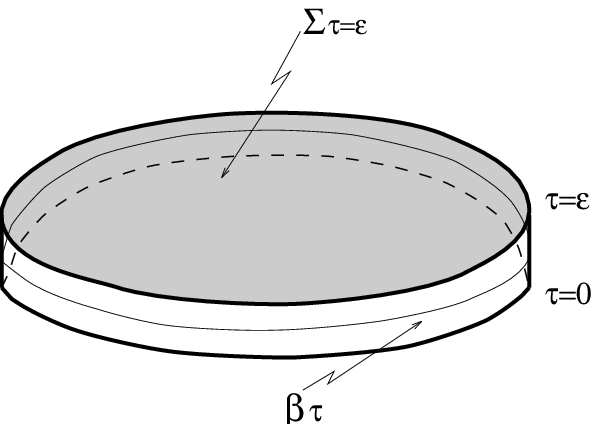}}\hfill}
\bigskip

\noindent 
{\small {\bf Fig. 3} The surface $S_\beta$ with boundary $\beta$ is
thickened to a three dimensional pill-box $P_\beta$ bounded by the
strip $\Sigma_\beta$. As $\epsilon$ tends to zero, $P_\beta$ shrinks
to $S_\beta$ and $\Sigma_\beta$ tends to the loop $\beta$. The
`memory' of the strip is retained by the initial framing attached to
$\beta$.}
\bigskip

We can now consider the flux of the electric field through the {\it
three dimensional} pill-box region
\be 
E[P_\beta] = \int_0^\epsilon \!\d\tau \, \int_{S_\tau}\! \d^2S_a\, 
E^a (\s_\tau) .
\ee
(From now on, loops will be assumed to be framed. However, for
simplicity of notation, we will continue to denote them just by greek
letters $\alpha, \beta, ..$.)  This yields a smearing of the electric
field with a test field $f^{(P_\beta)}_a$ with support in {\it
three}-dimensions,
\be
E[P_\beta] = \int\! \d^3x\, E^a(x) f^{(P_\beta)}_a(x) , 
\quad {\rm with}\quad
f_a^{(P_\beta)} = \int_0^\epsilon \d\tau\, \int_{S_{\tau}} (\d
S_{\tau})_a \,\delta^3(\x, \s_{\tau}) \, ,\label{covec}
\ee
and one can hope that the corresponding operator would be well-defined
in quantum theory.  This completes our first step in regularization.

The key question now is the following: Is $V = ({}^T\!f^{(P_\beta)}_a,
0)$ normalizable with respect to the inner product (\ref{norm})?  This
calculation is carried out in the Appendix. It turns out that,
although the 3-dimensional smearing softens the singularity of
$f^{(\beta)}_a$ considerably, the norm $\langle\!  V|V\!\rangle$ still
has a logarithmic ultra-violent divergence. This arises because the
three dimensional pill-box $P_\beta$, on which $f^{(P_\beta)}_a$ is
supported, has sharp boundaries. This is where the second step in the
regularization procedure comes in.  The problem can be handled in a
number of ways. We will use the simplest one and just introduce an
ultra-violet cut-off at $|{\bar k}|= \Lambda$. That is, we begin with
$f^{(P_\beta)}_a$ as in (\ref{covec}), take the Fourier transform of
its transverse part, multiply it by the step function which is unity
if $|{\bar k}| \le \Lambda$ and zero otherwise and consider the
inverse Fourier transform $f^{(P_\beta,\Lambda)}_a$ of the resulting
function. Then, $\hat{E}[f^{(P_\beta, \Lambda)}]$ is a well-defined
operator on the Fock space. This operator can be regarded as the
regulated version of the heuristic expression $\hat{E}[\beta]$ since,
in the classical theory, we have:
\be
\lim_{\epsilon \rightarrow 0} \frac{1}{\epsilon}\Big(
\lim_{\Lambda\rightarrow \infty}
\int \d^3x E^a(x) f^{(P_\beta, \Lambda)}_a (x)\,\Big) \, = E[\beta] 
\ee

The situation with the magnetic-flux operator, of course, is
identical.  Thus, given a framed loop $\alpha$, we can introduce a
strip $\Sigma_\alpha$ of height $\epsilon$ and consider a pill-box
$P_\alpha$ it bounds. Then, $\hat{B}[f^{(P_\alpha, \Lambda)}]$ is a
well-defined operator on the Fock space.

\section{Removal of the regulators}

Let us compute the commutator between the regularized flux operators
using (\ref{scr}) and then, in the result, remove the regulators by
taking appropriate limits. Removing the ultra-violet cut-off yields:
\ba
\lim_{\Lambda \rightarrow \infty} \Big[\hat{B}
[f^{(P_\alpha,\Lambda)}],\,\,
\hat{E}[f^{(P_\beta,\Lambda)]}\Big] &=& \hbar 
\lim_{\Lambda\rightarrow \infty}
\int_\Lambda \d^3k\, \epsilon^{abc}k_a\, \big(\tilde{f}^{(P_\alpha)}
(\k)\big)^\star\, \tilde{f}^{(P_\beta)}(\k)\nonumber\\
&=& i\hbar \int\!\d^3x\, \epsilon^{abc}\, \big(\partial_a
f_b^{(P_\alpha)}(\x)\big) f_c^{(P_\beta)}(\x)\nonumber\\ 
&=& i\hbar\, \int_0^\epsilon\d\sigma\,\oint_{\alpha_\sigma} \d l^a\,
\int_0^\epsilon \d\tau \, f_a^{(\beta_\tau)}\nonumber\\
&=& i\hbar\, \int_0^\epsilon \d\sigma \int_0^\epsilon \d\tau
\{ B[\alpha_\sigma],\,\, E[\beta_\tau] \}
\ea
where, in the first step, the subscript $\Lambda$ denotes that the
integration is carried out over the ball $|K| < \Lambda$. The final
result is not surprising: the right side is just $i\hbar$ times the
well-defined Poisson bracket between the classical `thickened flux'
observables:
\be
\lim_{\Lambda \rightarrow \infty} \Big[\hat{B}[f^{(P_\alpha,
\Lambda)}],\,\,
\hat{E}[f^{(P_\beta,\Lambda)}\Big] = i\hbar\, \{ B[f^{(P_\alpha)}],
\,\, E[f^{(P_\beta)}]\} . 
\ee
Thus, as far as the commutator is concerned, the ultra-violet cut-off
plays no essential role.

To remove the regulator $\epsilon$, we have to compute:
\ba
\lim_{\epsilon\rightarrow 0}\, \frac{1}{\epsilon^2}\,\Big(
\lim_{\Lambda \rightarrow \infty} \Big[\hat{B}[f^{(P_\alpha,\Lambda)}],
\,\,\hat{E}[f^{(P_\beta,\Lambda)}]\Big]\,\Big) &=& i\hbar\, 
\lim_{\epsilon\rightarrow 0}\, \frac{1}{\epsilon^2}\, \Big(
\int_0^\epsilon \!\d\sigma \int_0^\epsilon \!\d\tau\,
\{ B[\alpha_\sigma], E[\beta_\tau] \}\,\Big)\nonumber \\
&=& i\hbar \lim_{\epsilon\rightarrow 0}\, \frac{1}{\epsilon^2}\, 
\Big(\int_0^\epsilon\!\d\sigma \int_0^\epsilon\! \d\tau\,
{\cal P}(\alpha_\sigma,\, \beta_\tau)\Big)\, ,\,\,\, {\rm say.}
\ea
This calculation is more subtle. We will divide the discussion in four
cases which bring out the role of framing in handling pathologies that
arise when the loops intersect and overlap.

i) The simplest case arises when the loops have no point in common.
Then, for a sufficiently small $\epsilon$, there is no intersection
between any of the loops $\alpha_\sigma$ and $\beta_\tau$. Hence, the
Poisson bracket ${\cal P} (\alpha_\sigma,\, \beta_\tau)$ can be
calculated exactly as in Sec. I. It is independent of $\sigma$ and
$\tau$ and equals $\GL (\alpha, \beta)$. Hence,
\be
\lim_{\epsilon\rightarrow 0}\, \frac{1}{\epsilon^2}\, 
\int_0^\epsilon\!\d\sigma \int_0^\epsilon\! \d\tau\,
{\cal P}(\alpha_\sigma, \beta_\tau) = \GL(\alpha, \beta).
\ee
Thus, in this case, the limiting procedure gives a precise meaning to
the calculation of Sec. I: The uncertainty relation (\ref{ur})
holds in the sense that:
\be
\lim_{\epsilon\rightarrow 0}\, \frac{1}{\epsilon^2}\,\Big(
\lim_{\Lambda \rightarrow \infty} \Big[\hat{B}[f^{(P_\alpha,\Lambda)}],
\,\hat{E}[f^{(P_\beta,\Lambda)}]\,\Big]\Big) = i\hbar \GL(\alpha,\beta)
\ee

ii) Let us now consider the case when $\alpha$ and $\beta$ intersect
at a single point, say $p$. (Intersections at a finite number of
points requires only a trivial extension of this case.) Now, the
result depends on the thickening, or more precisely, on the framing at
$p$ initially chosen to carry out the thickening. Let $\dot\alpha^a$
and $\dot\beta^a$ denote the tangent vectors to the two loops at
$p$. Consider the two dimensional plane they span in the tangent space
of $p$.  Suppose that the frame vectors of the loops $\alpha$ and
$\beta$ lie on opposite sides of the plane. Then (for sufficiently
small $\epsilon$) among loops $\alpha_\sigma$ and $\beta_\tau$, the
only ones which intersect are $\alpha_0$ and $\beta_0$, the original
loops.  Hence, ${\cal P}(\alpha_\sigma, \beta_\tau)$ is well-defined
in the $\sigma$, $\tau$ space except at the single point, $(\sigma =0,
\tau =0)$, which is of measure zero. Furthermore, at all other points,
${\cal P}(\alpha_\sigma, \beta_\tau)$ is independent of $\sigma$ and
$\tau$. Its value is precisely the Gauss linking number
$\GL(\alpha,\beta') = \GL(\alpha', \beta)$, where the primed loops are
obtained by moving the unprimed ones slightly along the framing
vectors. Thus, we have:
\be
\lim_{\epsilon\rightarrow 0}\, \frac{1}{\epsilon^2}\,\Big(
\lim_{\Lambda \rightarrow \infty} \Big[\hat{B}[f^{(P_\alpha,
\Lambda)}]\, ,\hat{E}[f^{(P_\beta,\Lambda)}]\,\Big]\Big) = 
i\hbar \GL(\alpha, \beta') = i\hbar \GL(\alpha', \beta)\, ,
\ee
which, in this case, is ($i\hbar$ times) the natural Gauss linking
number associated with the {\it framed} loops.

iii) Let us now consider the case where the two loops intersect at a
single point $p$ as before but where frame vectors at $p$ (lie on the
same side of the plane spanned by the two tangents and) are parallel.
Then the two strips $\Sigma_\alpha$ and $\Sigma_\beta$ intersect in a
line rather than a single point.  In this case the limit is more
delicate. A loop $\alpha_\sigma$ on $\Sigma_\alpha$ intersects a loop
$\beta_\tau$ on $\Sigma_\beta$ if and only if $\sigma =\tau$. Thus,
the calculation of Sec. I  for computing ${\cal P}(\alpha_\sigma,
\beta_\tau)$ goes through for the entire region of the parameter space
$(\sigma, \tau) \in [0,\epsilon]\times[0, \epsilon]$ except for the
diagonal. Again, since the diagonal is a set of measure zero, we can
ignore it. However, now the integrand ${\cal P}(\alpha_\sigma,
\beta_\tau)$ is no longer a constant on the entire parameter space. 
As a simple geometric picture reveals, it takes one value,
$\GL(\alpha, \beta')$, on one side of the diagonal and another value,
$\GL(\alpha', \beta)$, on the other, where as before the primed loops
are obtained by displacing the unprimed loops slightly in the
direction of framing. Hence, we now have:
\be
\lim_{\epsilon\rightarrow 0}\, \frac{1}{\epsilon^2}\,\Big(
\lim_{\Lambda \rightarrow \infty} \Big[\hat{B}[f^{(P_\alpha,
\Lambda)}]\, ,\hat{E}[f^{(P_\beta,\Lambda)}]\,\Big]\Big) = 
\frac{i\hbar}{2}\Big(\GL(\alpha, \beta') + \GL(\alpha', 
\beta)\Big)\, . 
\ee
The right side is ($i\hbar$ times) the average of the two possible
linking numbers one can obtain by displacing the loops $\alpha$ and
$\beta$ infinitesimally using the assigned framing, i.e., the
`natural' (extension of the) Gauss linking number associated with the
two given {\it framed} loops.

iv) Finally, let us consider the commutator between fluxes of electric
and magnetic associated with the {\it same} framed loop $\alpha$. In
this case the strips $\Sigma_\alpha$ and $\Sigma_\beta$ as well as the
`pill-boxes' $P_\alpha$ and $P_\beta$ coincide. Now, if $\sigma\not= 
\tau$ the loops $\alpha_\sigma$ and $\beta_\tau$ have no points in common.
Hence, the integrand ${\cal P}(\alpha_\sigma, \beta_\tau)$ is
well-defined everywhere except along the diagonal. However, in this
case, (outside the diagonal which we can ignore) the value of the
integrand is in fact constant, namely, the Gauss linking number
$\GL(\alpha, \alpha')$, where $\alpha'$ is again obtained by
displacing $\alpha$ slightly along the framing. Thus, in this case, we
have:
\be
\lim_{\epsilon\rightarrow 0}\, \frac{1}{\epsilon^2}\,\Big(
\lim_{\Lambda \rightarrow \infty} \Big[\hat{B}[f^{(P_\alpha,
\Lambda)}]\, , \hat{E}[f^{(P_\alpha,\Lambda)}]\, \Big]\Big) 
= i\hbar\, \GL(\alpha, \alpha')\, .
\ee
The right side is precisely the {\it self-linking number} of the
framed loop $\alpha$ \cite{5}. Note in particular that if the framing
is trivial (e.g., if all the frame vectors are parallel in the three
dimensional Euclidean space), the right side vanishes (even before
taking the limit $\epsilon\rightarrow 0$) . Thus, in this case, one
can simultaneously measure the fluxes of electric and magnetic fields
with {\it arbitrary} accuracy.

There are of course other cases one can analyze. For example, one can
consider loops with isolated intersections, where, however, framing at
the intersection points is not of the type considered in cases ii) and
iii) above. Given the two framings, the calculation of the limit of
the commutator is generally straightforward. Since the regularization
is geometric and the final limiting procedure refers only to
properties of ribbons obtained from framing, it is natural to interpret
the result as the Gauss linking number of framed loops in those cases
as well.

\section{Discussion}

In this note we have pointed out that there is a remarkable relation
between the Gauss linking number, the simplest link invariant, and the
Heisenberg uncertainty between fluxes of electric and magnetic fields,
the basic observables of the quantum Maxwell theory.  This uncertainty
is intrinsic in that it arises because of the fundamental quantum
fluctuations and persists even in the vacuum state. 

The precise sense in which this relation holds is rather subtle
especially if the loops in question intersect or overlap. In the
classical theory, given any closed loop $\alpha$, we can compute the
fluxes $B[\alpha]$ and $E[\alpha]$ of magnetic and electric fields
through any surface $S_\alpha$ bounded by the loop $\alpha$. To obtain
the corresponding quantum observables, however, we have to `thicken'
the surfaces in question. A natural strategy is to frame the initial
loop since a framing provides a canonical thickening. When this is
done (and an ultra-violet cut-off is introduced) one obtains regulated
flux operators which are well-defined on the Fock space. We can
compute their commutators and {\it then} remove the regulators. The
limit is the just ($i\hbar$ times) the Gauss link invariant of the
framed loops. Even when the loops intersect or coincide (as in cases
ii), iii) and iv) considered in Sec. III), the limit of the commutator
equals the Gauss linking number of the {\it framed} loops.

For simplicity, we worked in Minkowski space-time. However, the entire
discussion can be carried over without any difficulty to general
stationary space-times (where the norm of the Killing field is bounded
away from zero). In this case, one can use geodesics tangential to the
framing to thicken the the loops and work in the canonical Fock
representation selected by the Killing field \cite{6,7}. In the
non-stationary context, there is no canonical representation of the
CCR.  However, one can again construct the algebra of smeared flux
operators and the basic results will hold on any Hilbert space on
which this algebra can be represented. This is to be expected because
the final results are topological and do not refer to the Minkowskian
geometry used in the intermediate stages.

Finally, we wish to point out that the Gauss linking number also plays
a key role in the expression of the measure which dictates the inner
product on the photon Hilbert space in the so-called self-dual
representation (where states are appropriate functionals of self-dual
connections) \cite{8}.

\section*{Acknowledgments}

We would like to thank Jorge Pullin for discussions.  This work was
supported in part by the NSF grant PHY95-14240 and by the the Eberly
Research Fund of Penn State University. In addition, AA received
partial support from the Erwin Schr\"odinger International Institute
for Mathematical Sciences, Vienna, and AC from DGAPA of UNAM in
Mexico.

\begin{appendix}
\section*{}

\subsection{From Intersection to Gauss-linking Number}

In this sub-section we shall show the analytic equivalence between the
intersection number $I(\alpha,S_\beta)$ of Eq (\ref{ino}) and the
Gauss linking number of Eqs (\ref{g},\ref{glno}).  We start by
re-writing the intersection number,
\ba
I(\alpha,S_\beta) &=& \{B[\alpha], E[\beta]\}\nl &=& \int \d^3x\,
F^a[\alpha,\x)w_a[\beta,\x)
\ea
where $F^a[\alpha,\x)$ is the so-called {\it form factor} of the loop
$\alpha$:
\be
F^a[\alpha,\x)=\oint_\alpha\d \alpha^a\delta^3(\alpha,\x)
\ee
and $w_a[\beta,\x)$ is given by
\be
w_a[\beta,\x)=\int_{S_\beta}\d S_a \,\delta^3(\x ,\s_\beta)
\ee
Also, note that $w_a$ is a potential for the form factor, since,
\be
F^a[\alpha,\x)=\epsilon^{abc}\partial_b w_c[\alpha,\x)
\ee
There is an extra `gauge freedom' since $w_a$ and $w^\prime_a=w_a
+\partial_a f$ give rise to the {\it same} form factor $F^a$.

So far, the intersection number does not depend on any background
structure and is therefore, topological in nature. The expression
(\ref{g}) is, however, written in terms of an Euclidean metric. Let us
therefore introduce such a metric. Then,
\be
\delta^3(\x-\y)=-\frac{1}{4\pi} \nabla^2_x\,\frac{1}{|\x-\y|}=
-\frac{1}{4\pi}
\partial_a\partial_x^a\left(\frac{1}{|\x-\y|}\right)\, .
\ee
We can now re-write the intersection number as,
\ba
I(\alpha,S_\beta) &=&-\frac{1}{4\pi}\int \d^3x\int\d^3y\,
F^a[\alpha,\x)\, w_a[\beta,\y)\,\partial_d\partial^d_y
\left(\frac{1}{|\x-\y|}\right)\nl
&=& \frac{1}{4\pi}\int\d^3x\int\d^3y\,
F^a[\alpha,\x)\,\partial_dw_a[\beta,\y)\,\partial^d_y
\left(\frac{1}{|\x-\y|}\right)
\ea
where in the second step we have integrated by parts. Now,
\be
I(\alpha,S_\beta) = \frac{1}{4\pi}\int\d^3xF^a[\alpha,\x) \int\d^3y\,
\left( 2\partial_{[d}w_{a]}[\beta,\y)+(\partial_a w_d[\beta,\y))
\partial^d_y\left(\frac{1}{|\x-\y|}\right)\right)
\ee
The last term can be again integrated by parts
\be
\int\d^3y (\partial_a w_d[\beta,\y))
\partial^d_y\left(\frac{1}{|\x-\y|}\right)=-\int\d^3y\frac{1}{|\x-\y|}
(\partial_a\partial^d 
w_d[\beta,\y))=0\, ,
\ee
where we have used the gauge freedom to select $w_a$ such that 
$\partial^a w_a=0$.

Finally, we have,
\ba
I(\alpha,S_\beta)&=& \frac{1}{4\pi} \int\d^3x\int\d^3y \,F^a[\alpha,\x)F^b
[\beta,\y) \epsilon_{abc}\partial_y^c\left(\frac{1}{|\x-\y|}\right)\nl
&=& \frac{1}{4\pi}\int\d^3x\int\d^3y \,F^a[\alpha,\x)F^b[\beta,\y) 
\epsilon_{abc} \frac{(x^c-y^c)}{|\x-\y|^3}\, .
\ea
Now, the definition of the form factor implies that: $\int\d^3x
F^a[\alpha,\x)f_a=\oint_\alpha\d\alpha^a f_a$. Hence, we have the desired
equality:
\be 
I(\alpha, S_\beta) = 
\frac{1}{4\pi}\,\, \oint _{\alpha}\d s \oint_{\beta}
\d t\,\,\epsilon_{abc} \,\dot{\alpha}^a(s)\,
\dot{\beta}^b(t) \frac{\alpha^c(s)- \beta^c(t)}{|\alpha(s) - \beta(t)|^3}\, .
\ee

\subsection{Quantum Operators}

Recall from Sec. II that the smeared operators
$\hat{E}[f^{(P_\beta)}]$ are well-defined on the Fock space $\F$ if
and only if the $\H$-norm of $V = ({}^T\!\!  f_a^{(P_\beta)}, 0)$ is
finite. In this sub-section we will compute this norm and show that it
has a logarithmic divergence thereby establishing the necessity of an
ultra-violet cut-off.

Let us consider the simplest case. In cylindrical coordinates $(\rho,
\phi, z)$, let the strip $\Sigma_\beta$ defining the pill-box region
$P_\beta$ be circles of radius $\rho =R$ and $z=\tau$; thus
$\vec{\alpha}_\tau(s)=(R, 2\pi \,s,-\epsilon /2 + \tau)$.  We can take
the $S_\tau$ surfaces to be parallel to the $z={\rm const}$ plane.  In
this geometry, the smearing co-vector field $f_a(x)$ is the `step
function': $f_a(x)=\nabla_az$ if $\rho<R$ and $z
\in [-\epsilon /2,\epsilon /2]$; $f_a(x)=0$ otherwise.  The Fourier
transform will have non-vanishing component only in the $k_z$
direction:
\be
\tilde{f}_{k_z}(k) =\frac{1}{(2\pi)^{3/2}} \int^{R}_0\int^{2\pi}_0 
\int^{\epsilon /2}_{
\epsilon /2} \d z\,\d\phi
\,\rho\, \d\rho \, e^{i k_{\rho} \rho \cos \phi}\, e^{i k_z z}
\ee
Using the identity,
\be
{\rm J}_0(z)= \frac{1}{\pi} \int^\pi_0 e^{i z \cos \theta }\,\d\theta
\ee
and the recurrence formulae of Bessel functions we arrive at
\be
\tilde{f}_{k_z}(k) = \sqrt{\frac{2}{\pi}} R\, \frac{{\rm J}_1
 (k_{\rho} R)}{k_{\rho}} \,
\frac{\sin (k_z \epsilon)}{k_z}
\ee

The transverse part any field $\tilde{f}^{T}_{a}(\k)$ is the
projection of the that field orthogonal to the radial vector $k^a$.
Therefore,
\be
|\tilde{f}^{T}|^2=|\tilde{f}_{k_z}|^2\frac{k_{\rho}^2}{(k_{\rho}^2+k^2_z)}
\, .
\ee
The expression $\int_{\Sigma}\d^3k\,|k|\,|\tilde{f}^{T}_a(k)|^2 $ 
now takes the form
\be
\int_{\Sigma}\d^3k\,|k|\,|\tilde{f}_a(k)|^2 = 4 R^2
\int_{-\infty}^{\infty} \int_0^{\infty}
\,\d k_z\,\d k_{\rho} \frac{k_{\rho}^2}{\left[k_z^2 + k_{\rho}^2\right]^{1/2}} 
\frac{{\rm J}^2_1(\kappa R)}{k_{\rho}}\,\,\frac{\sin^2(k_z \epsilon)}
{k_z^2}
\ee
where $|k|=[k_z^2 + k_{\rho}^2]^{1/2}$. It is now obvious that the
integral diverges logarithmically since ${\rm J}_1(x)\sim x^{-1/2}$
when $x\rightarrow \infty$.

\end{appendix}


\begin{thebibliography}{19}

\bibitem{1} C. F. Gauss,  {\em Werke} vol~V (G\"otingen, 
K\"onigliche Gesellschaft der Wissenschaften, 1833) 605, Note of January 22.

\bibitem{2} J. C. Maxwell, {\em A Treatise on Electricity and 
Magnetism} (Oxford, Clarendon, England, 1873), p 40-42.

\bibitem{5} R. Gambini  and J. Pullin,  {\em Loops, Knots, Gauge 
Theories and Quantum Gravity} (Cambridge University Press, Cambridge, 
England, 1996).

\bibitem{3} A. Ashtekar, Mathematical problems of non-perturbative
quantum general relativity, in {\em Proceedings of the 1992 Les Houches
School on Gravitation and Quantizations} ed B. Julia and J. Zinn-Justin
(Elsevier, Amsterdam, 1995).

\bibitem{9} A. Ashtekar,  J. Lewandowski, D. Marolf, J. Mourao, T. Thiemann.
 {\em J. Math. Phys.} {\bf 36}, 6456 (1995).

\bibitem{4} I. Bialynicki-Birula  and Z. Bialynicka-Birula, {\em 
Quantum Electrodynamics} (Pergamon, Oxford, England, 1975).

\bibitem{6} A. Ashtekar and A. Magnon,  {\em Proc. R. Soc.}
(London) {\bf A46}, 375 (1975).

\bibitem{7} B. S. Kay,   {\em Commun. Math. Phys.} {\bf 71}, 29 (1980)

\bibitem{8} A. Ashtekar and A. Corichi, {\em Class. Quantum Grav.}
{\bf 14}, A43 (1997).


\end{thebibliography}
\end{document}